\documentclass[showpacs,aps,pra,floats,twocolumn]{revtex4-1}
 \pdfoutput=1
\usepackage{graphicx}
\usepackage{braket}
\usepackage{color}
\usepackage{mathtools}
\usepackage{amsmath}
\usepackage{amssymb}
\usepackage{hyperref}
\usepackage{float}
\graphicspath{{./Afig/}}
\begin{document}

\newcommand{\sigp}{\sigma^+}
\newcommand{\sigm}{\sigma^-}
\newcommand{\Gspph}{\Gamma^{\sigma^+}_0}
\newcommand{\Gsmph}{\Gamma^{\sigma^-}_0}
\newcommand{\Gcdph}{\Gamma^\mathrm{cd}_0}
\newcommand{\vc}[1]{{\boldsymbol{\mathrm{#1}}}}
\newcommand{\comment}[1]{}
\newcommand{\remove}[1]{}
\newcommand{\quot}[1]{\textquotedblleft{}#1\textquotedblright}
\newcommand{\un}{\mathrm}
\newcommand{\B}{\langle B \rangle}
\newcommand{\blue}[1]{{\color{blue}#1}}
\newcommand{\red}[1]{{\color{red}#1}}
\makeatletter
\newcommand{\vast}{\bBigg@{4}}
\newcommand{\Vast}{\bBigg@{5}}
\makeatother
\title{Influence of electron-phonon scattering  for an on-demand quantum dot single-photon source using cavity-assisted  adiabatic passage}
\author{Chris Gustin}
\email{c.gustin@queensu.ca}
\author{Stephen Hughes}
\affiliation{\hspace{-40pt}Department of Physics, Engineering Physics, and Astronomy, Queen's University, Kingston, Ontario K7L 3N6, Canada\hspace{-40pt}}
\date{\today}

\begin{abstract}
We study the role of electron-phonon scattering for a pulse-triggered quantum dot single-photon source which utilizes a modified version of stimulated Raman adiabatic passage and cavity-coupling. This on-demand source is coherently pumped with an optical pulse in the presence of a continuous wave laser drive, allowing for efficient generation of indistinguishable single photons with polarizations orthogonal to the applied fields. In contrast to previous studies, we explore the role of electron-phonon scattering on this semiconductor system by using a polaron master equation approach to model the biexciton-exciton cascade and  cavity mode coupling. In addition to background zero-phonon-line  decoherence processes, electron--acoustic-phonon coupling, which usually degrades the indistinguishability and efficiency of semiconductor photon sources, is rigorously taken into account. We study how cavity and laser detunings affect the device performance, and explore the effects of finite temperature on pure dephasing and intrinsic phonon-coupling. We describe how this biexciton-exciton cascade scheme allows for true single photons to be generated with over 90\% quantum indistinguishability and efficiency simultaneously using realistic experimental parameters. We also show how the double-field dressing can be probed through the cavity-emitted spectrum.
\end{abstract}
\maketitle

\section{Introduction}
Integral to many schemes of quantum information processing, including linear quantum computation~\cite{knill01} and quantum cryptography~\cite{hiskett06}, is a deterministic source of on-demand single-photons. Effective on-demand single-photon sources are efficient quantum light sources (emitting a single photon each time they are triggered) which produce photons that are indistinguishable in frequency, polarization, and bandwidth. In practical sources, the important figures-of-merit are degraded by decoherence arising from the source coupling to its environment, typically containing a large number of quantum degrees of freedom. Decoherence (particularly optical dephasing) has proven to be a substantial barrier to practical implementation of nanotechnology which incorporates quantum mechanical phenomena~\cite{kok07, nazir09, bylander03}. However, some of these  barriers are not necessarily fundamental ones, and methods exist and are improving which reduce (and manipulate) decoherence in single-photon sources~\cite{reed16, somaschi16,kuhlmann13}.

 Nanoscale semiconductor quantum dots (QDs) function as solid-state ``artificial atoms" and are among the most promising of candidates for scalable single-photon sources~\cite{buckley12,unitt05}. The presence of an electronic bandgap allows for an optically active transition between ground and spatially-confined excited electron-hole pair (exciton) states, mimicking a two (or more) level atom, but with notable advantages including longer stability, tunable transition frequency~\cite{lodahl15}, and ease of implementation in a solid-state environment. Quantum dot single-photon sources are often coupled to photonic environments, such as micropillar cavities~\cite{ulrich11} or photonic crystal defects~\cite{lodahl15}, to facilitate enhanced light-matter interaction and collection of emitted photons via cavity-quantum electrodynamics (cavity-QED), which also helps to minimize the coupling time to decoherence processes. Cavity coupling can be used to exploit coherent phenomena characteristic of the strong-coupling regime of cavity-QED, including coherent Rabi oscillations of excitonic populations---manifesting in strong field phenomena such as Autler-Townes splitting of exciton energy levels~\cite{xu07,reithmaier04},  Mollow triplet emission spectra~\cite{flagg09,ulhaq13,roy11,roy12}, or weak-coupling regimes  such as the Purcell enhancement of  spontaneous emission rates~\cite{gerard98, balet07}, allowing for increased collection efficiency of the emitted single-photons~\cite{ding16,somaschi16}. 

Semiconductor cavity-QD systems are some of the most promising candidates for efficient, deterministic sources of indistinguishable single photons~\cite{lodahl15}. However, QD systems are not without their drawbacks. In particular, the solid-state nature of the QD leads to coupling of excitons with phonons, most notably, with longitudinal acoustic (LA) 
phonons~\cite{nazir16,ramsay10,2ramsay10,krummheuer02,vagov02,forstner03,besombes01}. This coupling is often an intrinsic source of decoherence within QD photon sources and imposes fundamental limits on the efficacy of QDs as quantum light sources~\cite{ilessmith16}. In particular, incoherent excitation schemes to invert exciton or biexciton populations typically involve pumping (e.g., with an above-resonant optical pulse) the QD to a higher energy level in the conduction band, and letting the QD relax to the exciton state through phonon-mediated transitions or other non-radiative processes~\cite{glassl13,bounouar15,quilter15,toda99}. This can introduce a timing ``jitter", or uncertainty, in the lifetime of the exciton state, reducing the indistinguishability of photons emitted~\cite{kiraz04,kaer13,nazir09}. Additionally, off-resonant excitation requires higher pump strengths, increasing phonon-induced dephasing rates~\cite{ross16}. As a result, incoherent off-resonant excitation schemes can produce photons with high efficiency (and robust to laser detunings) but poor indistinguishability, and are typically inferior to coherent pumping mechanisms for single-photon sources~\cite{ross16,pathak10}. In contrast, this work studies a coherent pump-triggered excitation scheme via a modified version of stimulated Raman adiabatic passage (STIRAP) and cavity coupling. This scheme, initially proposed by Pathak and Hughes~\cite{pathak10} for a simple 4-level atom system, uses the biexciton-exciton cascade---consisting of two linearly polarized excitons and a biexciton (two excitons) state---to coherently generate on-demand single photons of different polarization than the input pulse, allowing for spectral separation of pump from output light. While this work shows promise, the original calculations were performed without any inclusion of electron-phonon coupling effects, which are known to play an important role on semiconductor QD-cavity systems~\cite{nazir16,ulrich11,roy12,kaer13,roychoudhury15, ilessmith16,krummheuer02,hohenester10}.

In this work, we expand upon the simple atomlike Lindblad master equation (ME) approach~\cite{pathak10} by introducing into the analysis an explicit model of electron-phonon interactions via a time-convolutionless polaron ME. Polaron MEs have been successfully used to explain a variety of phonon-related phenomena in QD systems, including Rabi frequency renormalization~\cite{mccutcheon10}, phonon-modified QD emission spectra of exciton and biexciton states~\cite{roy11,hargart16}, off-resonant phonon-assisted population inversion~\cite{ross16}, and phonon-modified Purcell enhancement of spontaneous emission rates~\cite{roychoudhury15}. This polaronic ME approach allows one to numerically calculate the relevant figures-of-merit for a single-photon source, including the efficiency (quantified roughly via the expectation value of the number of photons emitted into the cavity), and the quantum indistinguishability, obtained from the two-time correlation functions of the cavity mode operators. The polaron transform is a unitary transform which shifts the analysis of exciton-phonon interactions to a quasiparticle ``polaron" frame, where in the limit of zero laser or cavity coupling, the independent Boson model (IBM)~\cite{mahan} is recovered exactly, treating certain phonon coupling to a Fermionic atom nonperturbatively. This allows for  accurate and efficient computations to be made over a wide variety of temperatures, where other methods---such as a weak-coupling ME approach, which is perturbative in the exciton-phonon interaction and thus does not capture multiphonon processes~\cite{nazir16}---fail. The polaron ME approach also has certain benefits over numerically-exact path-integral approaches~\cite{vagov11} of allowing for more physical insight, as well as easier computation of quantum optics observables including two-time-correlation functions, which are necessary, e.g., for calculation of the single-photon indistinguishability.

In this work, we show how this QD-cavity scheme produces single-photons of simultaneously high efficiency and indistinguishability, and we explore the role of temperature and phonon coupling in detail.
 We find that this STIRAP set-up can allow for photons with simultaneously over 90\% efficiency and indistinguishability to be generated even in the presence of LA phonon coupling using resonant pulse excitation, and with near unity efficiency and over 80\% indistinguishability using off-resonant pulse excitation.
 The  layout of the rest of our paper is as follows: In Sec.~\ref{2} we describe the theoretical formalism of the open system time-dependent quantum dynamics we use to model the QD-cavity single-photon source. In Sec. \ref{3}, we solve the polaron ME for the system reduced density operator and two-time correlations of the cavity mode operators to calculate the expectation value of number of photons emitted into the cavity per pulse excitation, and quantum indistinguishability of emitted photons. We also calculate the cavity-emitted spectrum and explain its features in terms of the field-dressing of the system eigenstates by the various field and cavity couplings in the presence of phonon-coupling. Finally, we conclude in Sec.~\ref{conclusions}. 
\section{Theoretical Model}
\label{2}
\subsection{QD-Cavity System Hamiltonian}
\label{a}
\begin{figure}[!t]
\centering
\includegraphics[width=1\linewidth]{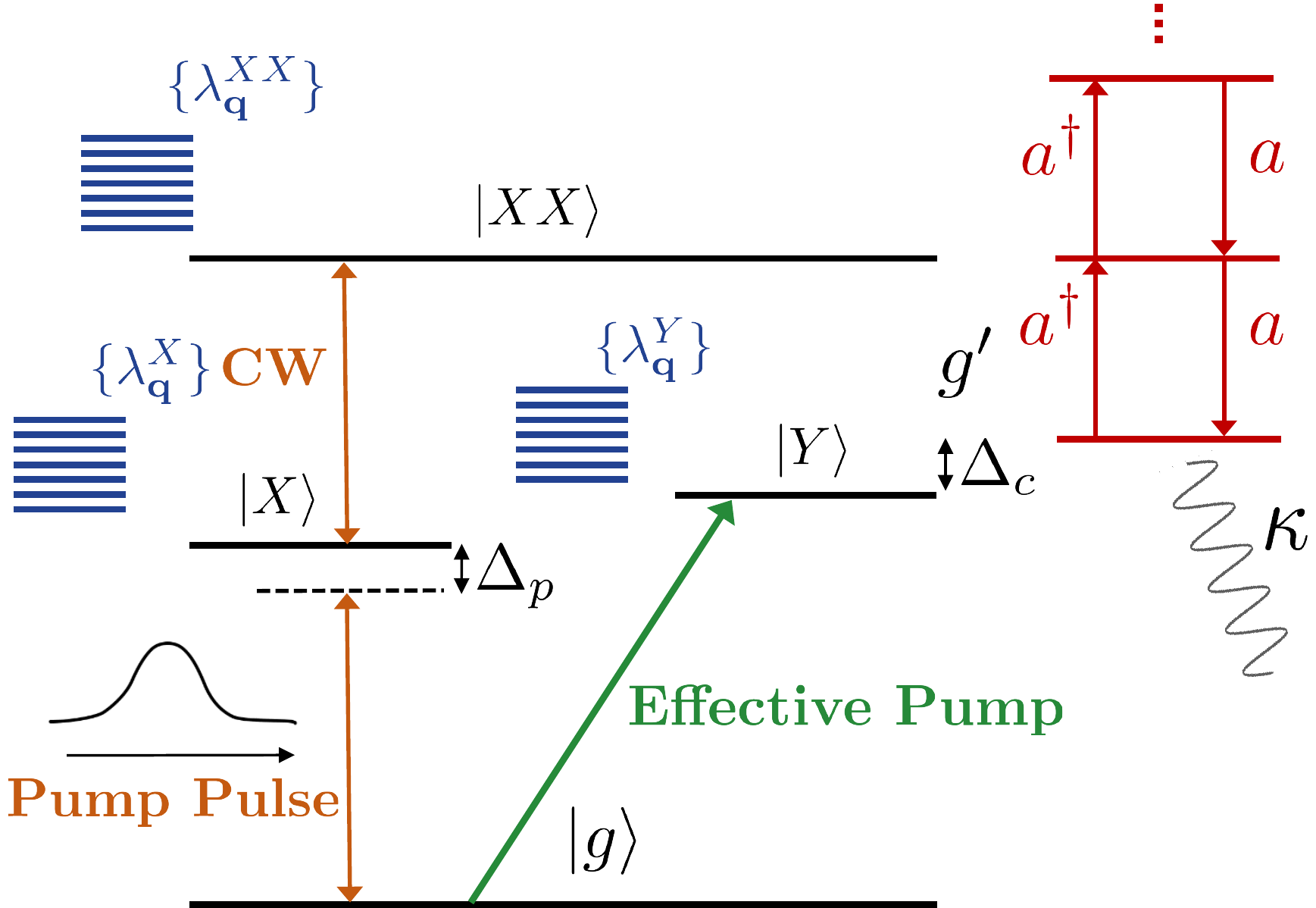}
\caption{\small Schematic of the STIRAP single-photon source excitation method in the biexciton cascade of a QD coupled to a cavity mode  and phonon bath.}
\label{stirap_fig}
\end{figure}

We model the QD-cavity system with a four-level biexciton cascade scheme (see Fig.~\ref{stirap_fig}) coupled to a cavity mode with creation and destruction operators $a^{\dagger}$ and $a$, respectively. Additionally, each excited state of the QD is coupled to a bath of phonon modes indexed by wavevector $\mathbf{q}$ and with bosonic creation and destruction operators $b^{\dagger}_{\mathbf{q}}$ and $b_{\mathbf{q}}$. The ground to $X$-exciton ($x$-polarized) transition is coupled by a time-dependent optical pump pulse with Rabi frequency $\Omega_ p(t)$, while the $X$-exciton to biexciton (XX) transition is coupled by a continuous wave (CW) laser with Rabi frequency $\Omega_l$. The cavity mode is chosen to couple the biexciton state to the $Y$-exciton with coupling constant $g$. With appropriate choice of drive strengths, the system population is adiabatically transferred via the STIRAP process from the ground state to the $Y$-exciton without  significantly populating the $X$-exciton and biexciton states, in the process emitting a single photon into the cavity. The system then decays radiatively to the ground state, allowing the source to be triggered once again. Neglecting, for now, background  decoherence of the zero-phonon line (ZPL) including cavity decay, spontaneous emission, and pure dephasing, the Hamiltonian for this system in a rotating frame (see Appendix for the relevant unitary transformation) is
\begingroup
\addtolength{\jot}{0.4em}
\begin{align}\label{ham}
H = & \ \hbar \Delta_p \ket{X}\bra{X} + \hbar(\Delta_p + \Delta_l - \Delta_c)\ket{Y}\bra{Y}  \nonumber \\  + & \ \hbar\big(\Omega_p(t)\ket{X}\bra{g} + \Omega_l\ket{XX}\bra{X} + g\ket{XX}\bra{Y}a + \text{H.c.}\big) \nonumber \\ + & \ \hbar(\Delta_p + \Delta_l)\ket{XX}\bra{XX} + \sum\limits_{\mathbf{q}}\hbar\omega_{\mathbf{q}}b_{\mathbf{q}}^{\dagger}b_{\mathbf{q}} \nonumber \\ + & \ \sum\limits_{\mathclap{S=\{X,Y,XX\}}} \ \ket{S}\bra{S}\sum\limits_{\mathbf{q}}\hbar\lambda_{\mathbf{q}}^{S}(b_{\mathbf{q}}^{\dagger}+b_{\mathbf{q}}),
\end{align}
\endgroup
with pump pulse detuning $\Delta_p \equiv \omega_X - \omega_p$, CW laser detuning $\Delta_l \equiv \omega_{XX} - \omega_l - \omega_X$, and cavity detuning $\Delta_c \equiv \omega_{XX} - \omega_Y - \omega_c$. For the multi-stage STIRAP process, the CW laser should be on resonance such that $\Delta_l = 0$, and $\Delta_p = \Delta_c$ to satisfy the multi-photon resonance condition~\cite{pathak10,vitanov99}. Thus in the following analysis we set $\Delta_l = 0$ and define $\Delta \equiv \Delta_p = \Delta_c$. To ensure that none of the transitions between ground and exciton states simultaneously couple to exciton to biexciton transitions (and vice-versa), the magnitude of the detuning $\Delta$ is assumed well below the biexciton binding energy (typically on the order of 1 meV) for the QD of interest such that $|\Delta| \ll |\omega_X - \frac{1}{2}\omega_{XX}|$. The LA phonon-exciton coupling is included via coupling constants $\{\lambda_{\mathbf{q}}^{S}\}$  for $S = \{X,Y,XX\}$, which are assumed to be real and correspond to an ideal quantum confined QD such that $\lambda_{\mathbf{q}} \equiv \lambda_{\mathbf{q}}^{X} = \lambda_{\mathbf{q}}^{Y} = \frac{1}{2}\lambda_{\mathbf{q}}^{XX}$~\cite{hohenester07}.
\subsection{Polaron master equation}
To analyze the dynamics of the STIRAP single-photon source, we develop an open-system time-local ME approach to determine the reduced system density matrix and calculate relevant quantities. To incorporate effects of the phonon bath, we first apply a polaron transform of the form $H' = e^{S}He^{-S}$ to approximately (exactly in the limit of no field couplings) diagonalize the Hamiltonian in the polaron frame, where $S = (2\ket{XX}\bra{XX} + \ket{X}\bra{X} + \ket{Y}\bra{Y})\sum_{\mathbf{q}}\frac{\lambda_{\mathbf{q}}}{\omega_{\mathbf{q}}}(b^{\dagger}_{\mathbf{q}}+b_{\mathbf{q}})$. Separating the polaron transformed Hamiltonian $H'$ into system, bath, and interaction parts such that $H' = H'_S + H'_B + H'_I$, we have
\begin{align}\label{hsys}
H&'_S =  (\Delta - \delta_P)\ket{X}\bra{X} \nonumber \\ + & (\Delta - 2\delta_P)\ket{XX}\bra{XX} - \delta_P\ket{Y}\bra{Y} \nonumber \\
+ & \big(\Omega'_p(t)\ket{X}\bra{g} + \Omega'_l\ket{XX}\bra{X}
 + g'\ket{XX}\bra{Y}a + \text{H.c.}\big),
\end{align}
where $\delta_P = \sum_{\mathbf{q}} \frac{\lambda_{\mathbf{q}}^2}{\omega_{\mathbf{q}}}$ is a  Lamb-shift in the exciton energies due to phonon bath renormalization. We can assume that this polaron shift is absorbed into the original definitions of $\Delta_p, \Delta_c$, and $\Delta_l$ such that it can be neglected henceforth. The drive strengths and cavity coupling $\Omega_p'(t)= \B\Omega_p(t), \Omega_l'=\B \Omega_l,$ and $g'=\B g$ are coherently reduced by the presence of the phonon bath, where $\B = \langle B_+ \rangle = \langle B_- \rangle 
= \text{exp}\Big[-\frac{1}{2}\sum_{\mathbf{q}}\frac{\lambda^2_{\mathbf{q}}}{\omega^2_{\mathbf{q}}}\coth{\big(\frac{\hbar\omega_{\mathbf{q}}}{2k_B T}\Big)}\Big]$ is the thermal average of the coherent displacement operators $ B_{\pm} = \text{exp}\big[\pm\sum_{\mathbf{q}}\frac{\lambda_{\mathbf{q}}}{\omega_{\mathbf{q}}}(b_{\mathbf{q}}^{\dagger}-b_{\mathbf{q}})\big]$. The bath free Hamiltonian is given by $H'_B = \sum_{\mathbf{q}}\hbar\omega_{\mathbf{q}}b_{\mathbf{q}}^{\dagger}b_{\mathbf{q}}$, and the interaction Hamiltonian is $H'_I = X_g \zeta_g + X_u \zeta_u$, with drive operators $X_g = \hbar\Omega_p(t)\ket{X}\bra{g} + \hbar\Omega_l\ket{XX}\bra{X} + \hbar g \ket{XX}\bra{Y}a + \text{H.c.}$, and $X_u = i\big(\hbar\Omega_p(t)\ket{X}\bra{g} + \hbar\Omega_l\ket{XX}\bra{X} + \hbar g \ket{XX}\bra{Y}a\big) + \text{H.c.}$, and phonon fluctuation operators $\zeta_g = \frac{1}{2}(B_+ + B_- - 2\B)$ and $\zeta_u = \frac{1}{2i}(B_+ - B_-)$.

We assume a continuous spectrum of phonon modes such that $J(\omega) = \sum_{\mathbf{q}}\lambda_{\mathbf{q}}^2\delta(\omega-\omega_{\mathbf{q}}) \rightarrow J(\omega) = \alpha \omega^3 \text{exp}\big[\frac{\omega^2}{2\omega_b^2}\big]$, which is the form of the phonon spectral function $J(\omega)$ appropriate for describing a deformation potential induced by LA phonons---the primary source of phonon-related decoherence in QD single-photon sources~\cite{nazir16,ramsay10,2ramsay10}; $\alpha$ is the exciton-phonon coupling strength, and $\omega_b$ is the phonon cut-off frequency. Following Refs.~\onlinecite{hargart16,roy12}, we derive a time-local 2nd-order Born-Markov ME in the polaron frame to treat the fluctuations in the phonon-exciton interaction induced by the optical fields perturbatively:
\begin{align}\label{me}
\frac{d}{dt}\rho(t) &= -\frac{i}{\hbar}[H'_S, \rho(t)] - \frac{1}{\hbar^2}\int_0^{\infty}d\tau \sum\limits_{\mathclap{m = \{g, u\}}}  \big(G_m(\tau) \nonumber \\ 
&\times [X_m(t),X_m(t,\tau) \rho(t)] + H.c. \big) + \sum_{\mu} \mathcal{L}[O_\mu]\rho(t),
\end{align}
where $G_g(\tau) = \B^2 (\cosh{(\phi(\tau))}-1)$ and $G_u(\tau) = \B^2\sinh{(\phi(\tau))}$ are the polaron Green functions, and $X_m(t,\tau) \approx e^{-iH'_S(t)\tau/\hbar} X_m(t) e^{iH'_S(t)\tau/\hbar}$. In the continuum limit, we have the following IBM phase function:
\begin{equation}
\phi(\tau) = \int\limits_{0}^{\infty}d\omega \frac{J(\omega)}{\omega^2}\bigg(\coth{\Big(\frac{ \hbar \omega}{2 k_B T}\Big)}\cos{(\omega\tau)} - i\sin{(\omega\tau)}\bigg),
\end{equation}
with $\B = e^{-\phi(0)/2}$, and this function includes multiple phonon absorption and emission transitions. Note that the polaron transform approach is rigorously valid in the regime where, for a given Rabi frequency, $\big(\frac{\Omega}{\omega_b}\big)^2(1-\B)^4 \ll 1$, which is the case for the parameters used here~\cite{mccutcheon10}. Additionally, to phenomenologically include decohering processes beyond LA phonon-exciton coupling~\cite{roy12}, we include Lindblad terms (for collapse operator $O$: $\mathcal{L}[O]\rho \equiv O\rho O^{\dagger} - \frac{1}{2}\{O^{\dagger}O, \rho\}$), corresponding to collapse operators  $\sqrt{\gamma_{XX}}\ket{X}\bra{XX}$, $\sqrt{\gamma_{XX}}\ket{Y}\bra{XX}$, $\sqrt{\gamma_{X}}\ket{g}\bra{X}$, $\sqrt{\gamma_{X}}\ket{g}\bra{Y}$ (i.e. $\gamma_{X} = \gamma_{Y}$) corresponding to spontaneous emission, as well as $\sqrt{2\gamma'}\ket{XX}\bra{XX}$, $\sqrt{\gamma'}\ket{X}\bra{X}$ and $\sqrt{\gamma'}\ket{Y}\bra{Y}$ corresponding to pure dephasing of exciton states. We furthermore introduce cavity photon leakage via collapse operator $\sqrt{\kappa}a$. Initially the QD (assumed to be neutrally charged) is taken to be in the ground state and the cavity mode to be in the vacuum state. 
\begin{figure}[bh]
\centering
\includegraphics[width=1\linewidth]{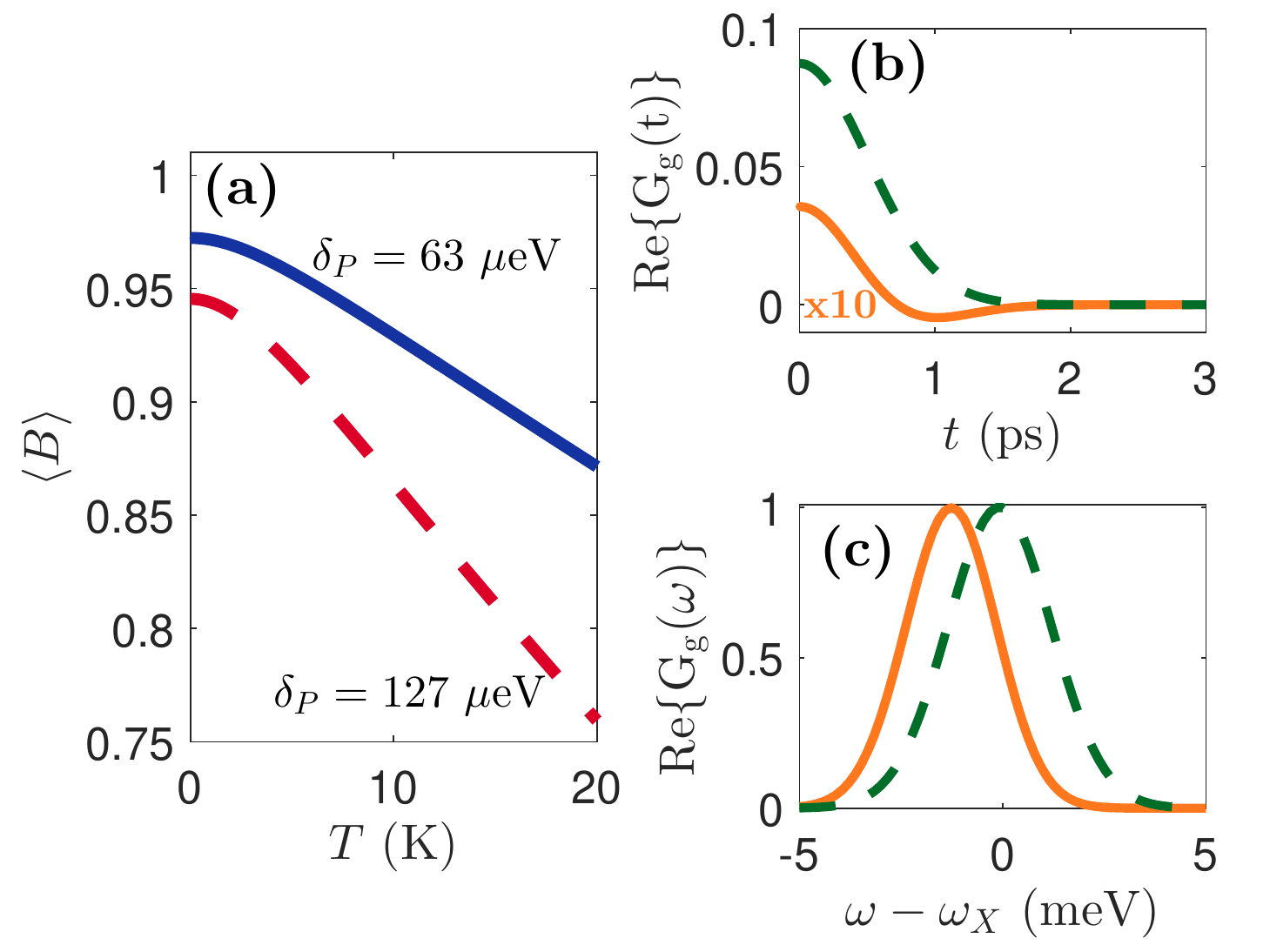}
\caption{\small (a) Phonon bath mean displacement $\B$ (drive strength and cavity coupling renormalization factor) as a function of temperature for $\alpha = 0.03 \ \text{ps}^2$ (blue, solid) and $\alpha = 0.06 \ \text{ps}^2$ (red, dotted), as well as the associated polaron shifts $\delta_P = \int_{0}^{\infty}d\omega J(\omega)/\omega$. (b) Real part of one of the polaron Green functions $G_g(t)$, giving the time evolution of the bath correlation function (for an exciton state) with $\alpha = 0.03 \ \text{ps}^2$ for $T = 5$ K (dotted green line) and $T = 40$ K (solid orange line). (c) Real part of $G_g(\omega)$ in the frequency domain, showing a low-temperature asymmetry of phonon bath correlations.}
\label{polaron_fig} 
\end{figure}

\begin{figure}[h]
\centering
\includegraphics[width=1\linewidth]{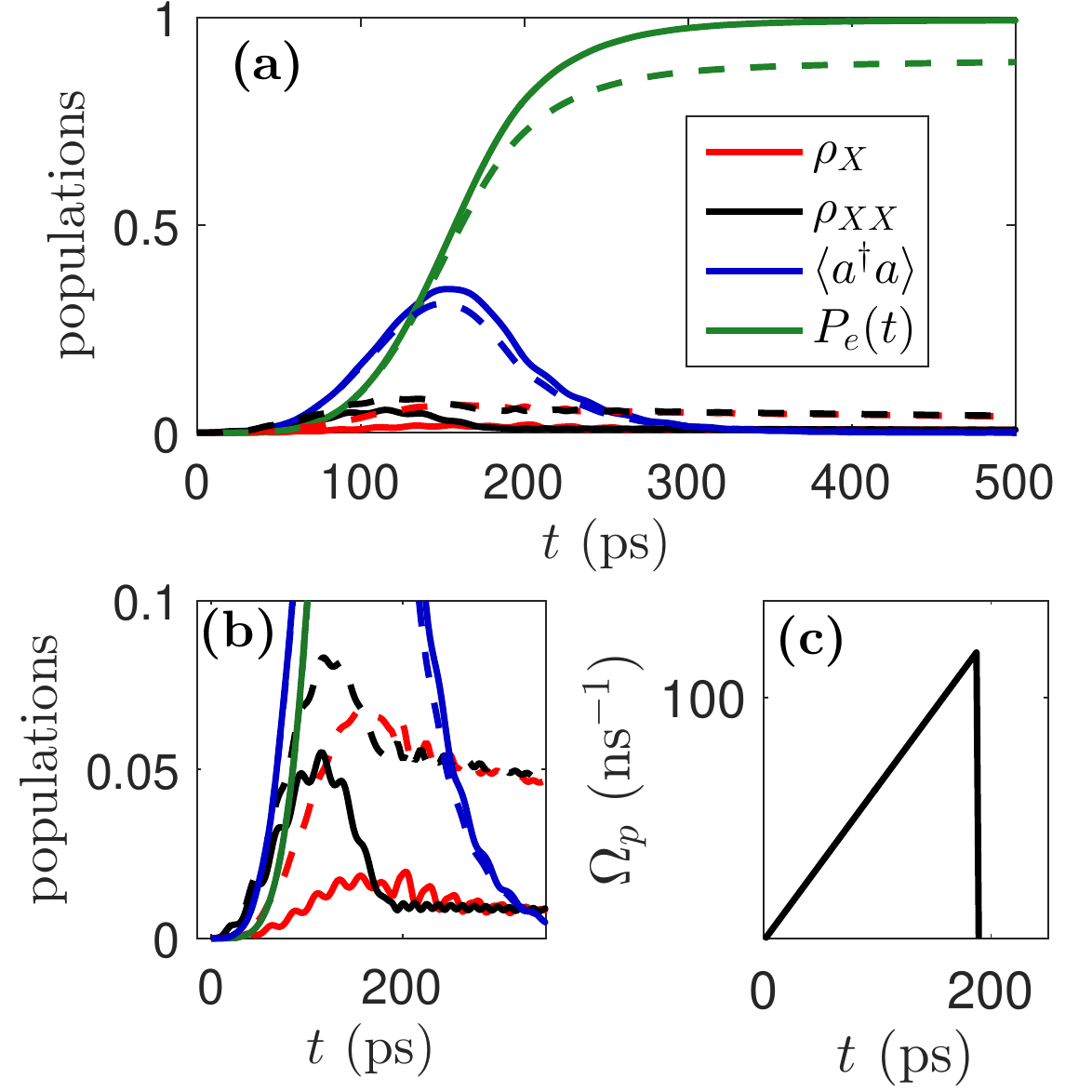}
\caption{\small (a) Populations of $X$-exciton ($\rho_X$; red), biexciton ($\rho_{XX}$; black), cavity ($\langle a^{\dagger} a \rangle$; blue), as well as the emitted photon number ($P_e(t)$; green) with (dashed lines) and without (solid lines) the exciton-phonon interaction. (b) Inset of plot in (a). (c) Pump pulse time profile, with max pulse strength $2.5g'$ and pulse width $3\pi/g'$.}
\label{Pop_fig} 
\end{figure}

\begin{figure}[h]
\centering
\includegraphics[width=0.95\linewidth]{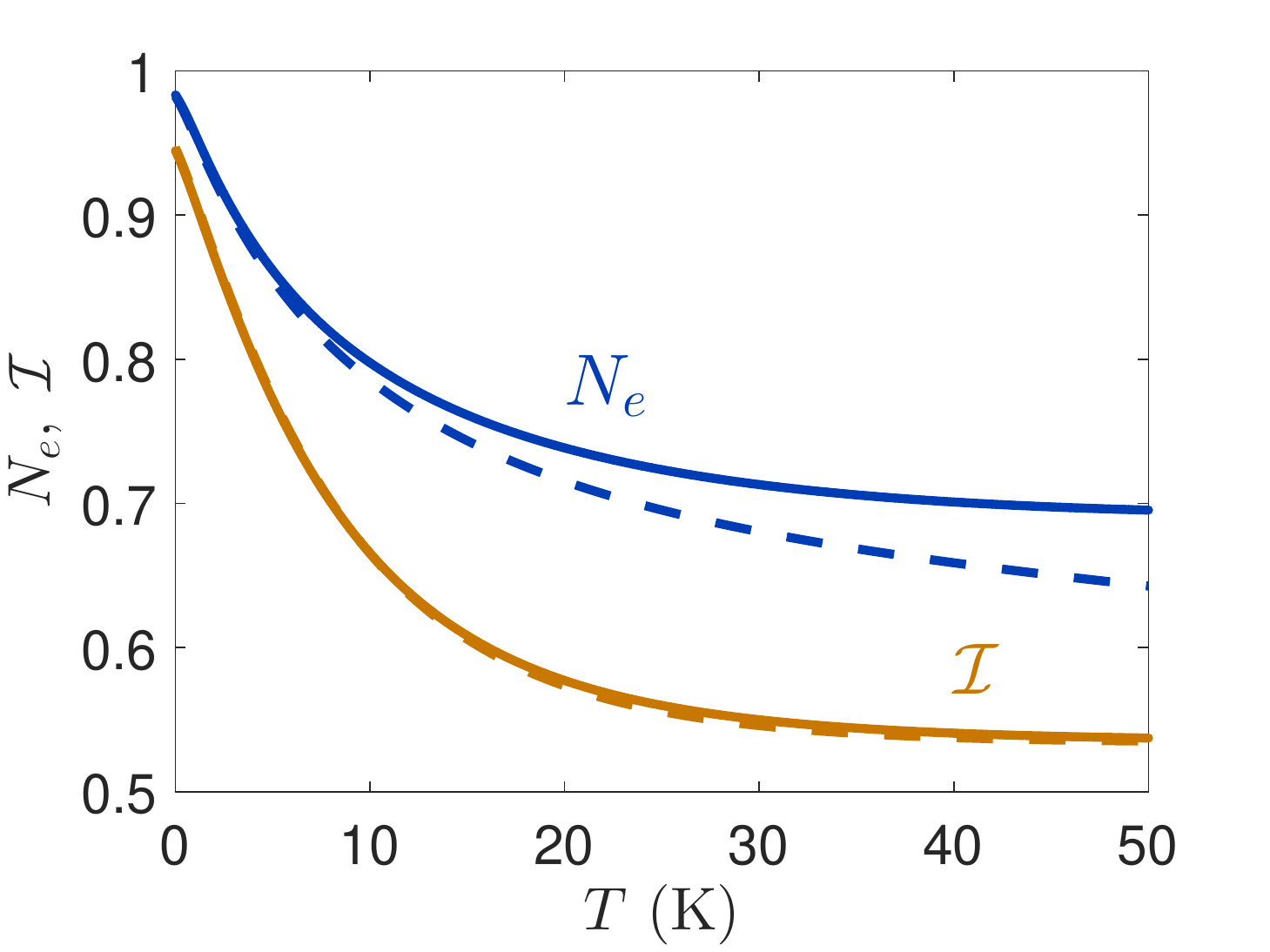}
\caption{\small Effect of phonon renormalization of excitation parameters. Shown is the emitted photon number and indistinguishability for constant (temperature-independent) bare coupling parameters $g$, $\Omega_l$, $\Omega_p$ (dashed lines) and constant effective coupling parameters $g'$, $\Omega_l$, $\Omega_p'$ (solid lines). We use a temperature-dependent dephasing rate $\gamma'(T) = \gamma'_0 +  (2.127 \ \text{ns}^{-1}\text{/K})T$ with $\gamma'_0 = 1 \ \text{ns}^{-1}$, as discussed in main text.}
\label{renorm_fig} 
\end{figure}

To calculate the key figures-of-merit for this single-photon source in the presence of phonons, the ME of Eq. \eqref{me} is first solved numerically. We then quantify the efficiency of the source with the emitted cavity photon number, $N_e \equiv \lim\limits_{t \rightarrow \infty} P_{e}(t)$, with $P_e(t) = \int_{0}^{t}\kappa\langle a^{\dagger} a \rangle(t') dt'$. Following previous works~\cite{pathak10,kiraz04,kiraz04E}, we can quantify the indistinguishability of the single-photons by considering a Hong-Ou-Mandel interferometry set-up, where two photons consecutively emitted from the single-photon source are directed at a beam splitter. The indistinguishability, ${\cal I}$ , of the photons determines the degree to which two-photon interference is observed, and can be expressed in terms of the cavity mode correlation functions:
\begin{equation}\label{indis}
\mathcal{I} \equiv \lim\limits_{T \rightarrow \infty}\frac{1}{2}\vast[1-\frac{\int\limits_{0}^{T}dt\int\limits_0^{T-t}d\tau\big[g^{(2)}(t,\tau)-|g^{(1)}(t,\tau)|^2\big]}{\int\limits_{0}^{T}dt\int\limits_0^{T-t}d\tau \langle a^{\dagger}a\rangle(t) \langle a^{\dagger}a\rangle(t+\tau)}\vast],
\end{equation}
where $g^{(1)}(t,\tau) = \langle a^{\dagger}(t)a(t+\tau) \rangle$ and $g^{(2)}(t,\tau) = \langle a^{\dagger}(t)a^{\dagger}(t+\tau)a(t+\tau)a(t)\rangle$ are the quantum degrees of first and second order coherence, respectively, which are calculated from the ME solution via the quantum regression theorem~\cite{carmichael}.
\section{Results}
\label{3}
\begin{figure}[t]
\centering
\includegraphics[width=1\linewidth]{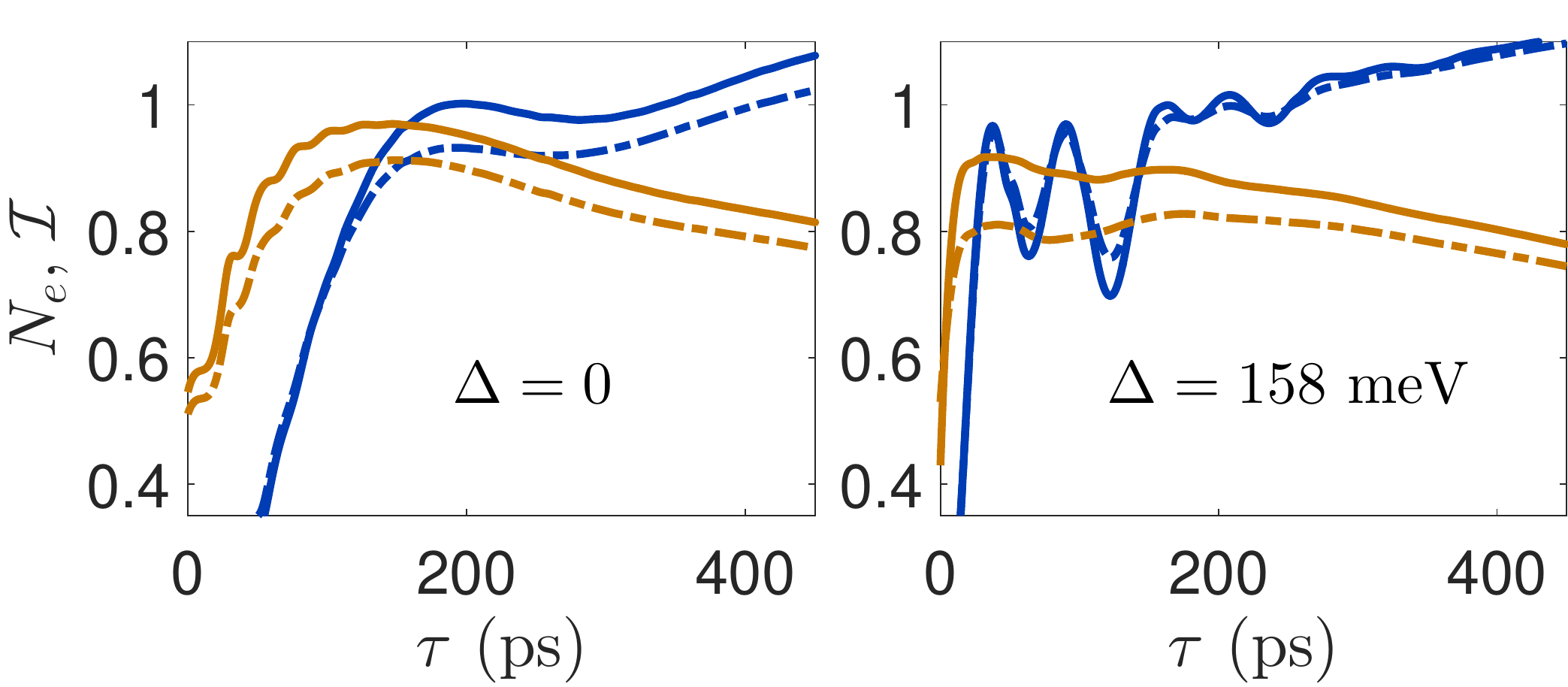}
\caption{\small Emitted cavity photon number (blue) and indistinguishability (brown) for on-resonant (left) and off-resonant (right) excitation with phonons (dashed lines) and without (solid lines).}
\label{pulsewidth_fig} 
\end{figure}

Throughout this work, we use parameters $\gamma_{XX} = \gamma_{X} = 0.5 \ \text{ns}^{-1}$ ($0.33 \ \mu \text{eV}$), and $\kappa = 25 \ \text{ns}^{-1}$ ($16.5 \ \mu \text{eV}$). Note that the value of $\kappa$ is important; if the cavity decay rate is too small, the cavity mode and biexciton state will sustain Rabi oscillations, and if it is too large, the biexciton state will decouple from the $Y$-exciton before the population transfer is complete, as $\ket{XX}$ does not coherently couple to the state $\ket{Y} \otimes \ket{0}$ (where $\ket{0}$ is the cavity mode vacuum state). The background pure dephasing rate (e.g., due to charge, spin noise), except where chosen as a parameter to vary, is $\gamma' = 1 \ \text{ns}^{-1}$ ($0.66 \ \mu \text{eV}$). The phonon parameters are chosen to be $\alpha = 0.03 \ \text{ps}^2$, and $\omega_b = 0.9$ meV, similar to those found from the experimental results by Quilter \emph{et al.}~\cite{quilter15}. The phonon coupling strength $\alpha$ can vary  from dot to dot, and so we also show in Fig.~\ref{polaron_fig}(a) the phonon bath mean displacement for a value of $\alpha = 0.06 \ \text{ps}^2$, as found in the experimental results by Weiler \emph{et al.}~\cite{weiler12}. To optimize the STIRAP process~\cite{kuhn02}, we take the pump pulse profile to be sawtooth (see Fig.~\ref{Pop_fig}) with max pulse strength $\Omega_p' = 2.5g'$ and pulse width $g'\tau_p = 3 \pi$ throughout, except in Fig.~\ref{pulsewidth_fig} where the pulse width is varied. The CW laser strength is fixed at $\Omega_l' = 5g'$. Except in Fig.~\ref{renorm_fig}, where we show the effects of the coherent renormalization of coupling parameters by the phonon bath, we take $g' = 50 \ \text{ns}^{-1}$ ($32.9 \ \mu \text{eV}$) to be a constant throughout, noting that the zero-temperature cavity coupling, Rabi frequencies, and pulse width must be modified by a factor of $\B$ to accurately compare the incoherent effects of the phonon bath, which are of interest, as they limit the efficiency and indistinguishability of the single-photon source. The renormalization effect is quite small at low temperatures, and a mean bath displacement of $\B = 0.96$ at $T = 5$ K only affects the emitted photon number by less than $0.005$ with these excitation parameters. Unless otherwise specified, all results with phonons are at  the bath temperature of $T=5$ K. Simulating results ``without phonons" means that the incoherent exciton-phonon scattering terms in Eq. \eqref{me} are set to zero and the cavity and field couplings are not renormalized by the factor of $\B$.

Figure~\ref{Pop_fig}  displays the populations of various QD states and the cavity mode over time, showing the influence of incoherent exciton-phonon scattering. Without phonons, these simulation parameters give an emitted photon number $N_e=1.00$ and indistinguishability $\mathcal{I}=0.96$, and with phonons, $N_e=0.93$ and $\mathcal{I}=0.90$. The presence of phonons indeed increases the degree to which the intermediate states ($X$-exciton and biexciton) in the STIRAP process are populated, decreasing the efficiency of the adiabatic population transfer. This can be attributed to additional phonon-induced dephasing captured in the polaron ME, reducing the coherence of the transfer process, as well as transitions between states mediated by phonon absorption and emission. Note that the finite lifetime of the $Y$-exciton state means that there is a small probability of the exciton decaying to the QD ground state and being re-excited during the same pump pulse, emitting two photons into the cavity. This limits the indistinguishability of the emitted photons and thus overly long pulse widths should be avoided. This also means that longer excited state lifetimes (specifically, the $Y$-exciton lifetime~\cite{pathak10}) improve the indistinguishability of the emitted photons, suggesting that this set-up could benefit from the reduction of the density of optical states away from cavity resonance, which, e.g., can be achieved with a photonic crystal cavity~\cite{roychoudhury15}.

\begin{figure}[!t]
\centering
\includegraphics[width=1\linewidth]{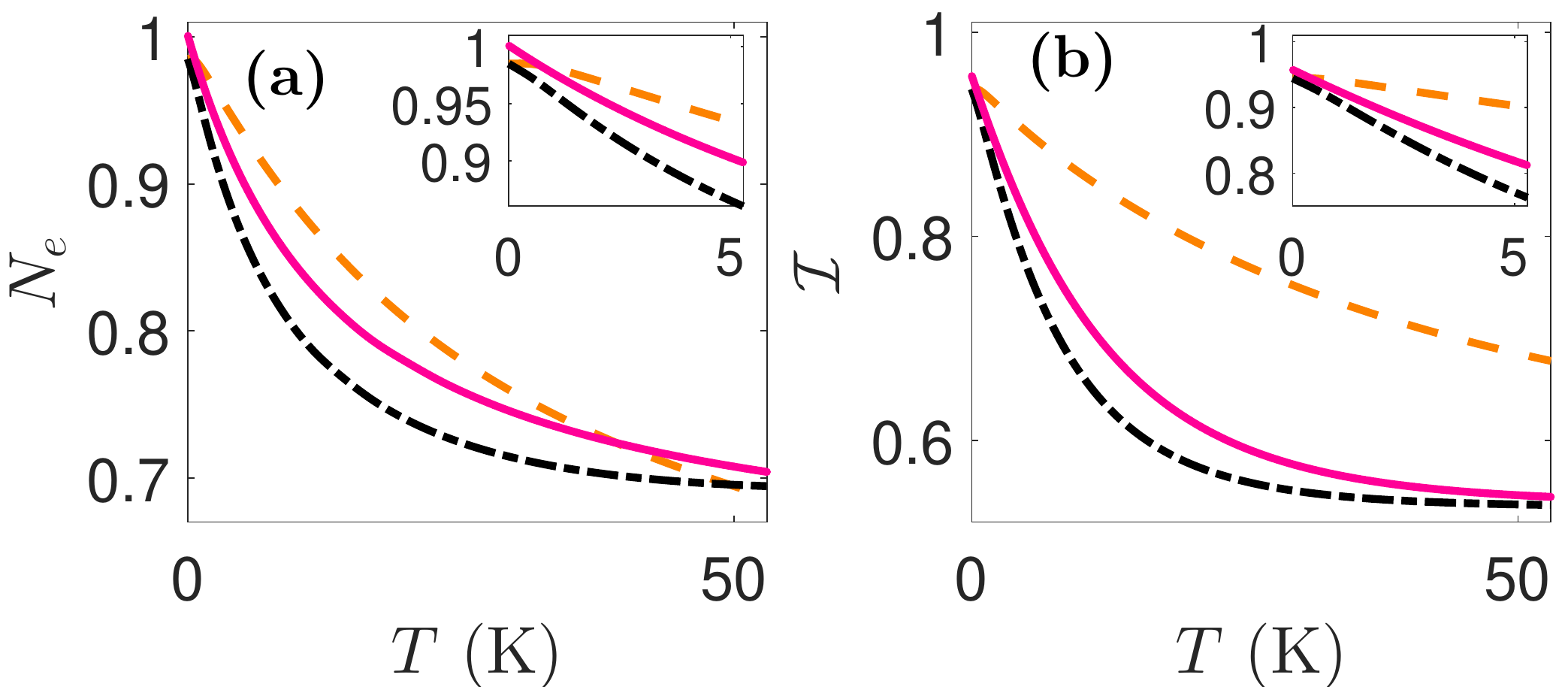}
\caption{\small (a) Number of photons emitted into cavity $N_e$ and (b) indistinguishability $\mathcal{I}$ as a function of temperature with phonons and constant dephasing $\gamma'_0 = 1 \ \text{ns}^{-1}$ (dashed orange line), with a temperature-dependent dephasing $\gamma'(T) = \gamma'_0 +  (2.127 \ \text{ns}^{-1}\text{/K})T$ and no phonons (solid magenta line), and with both phonons and temperature-variable dephasing (black dash-dotted line).}
\label{tempvar_fig} 
\end{figure}

\begin{figure}[bh]
\centering
\includegraphics[width=0.95\linewidth]{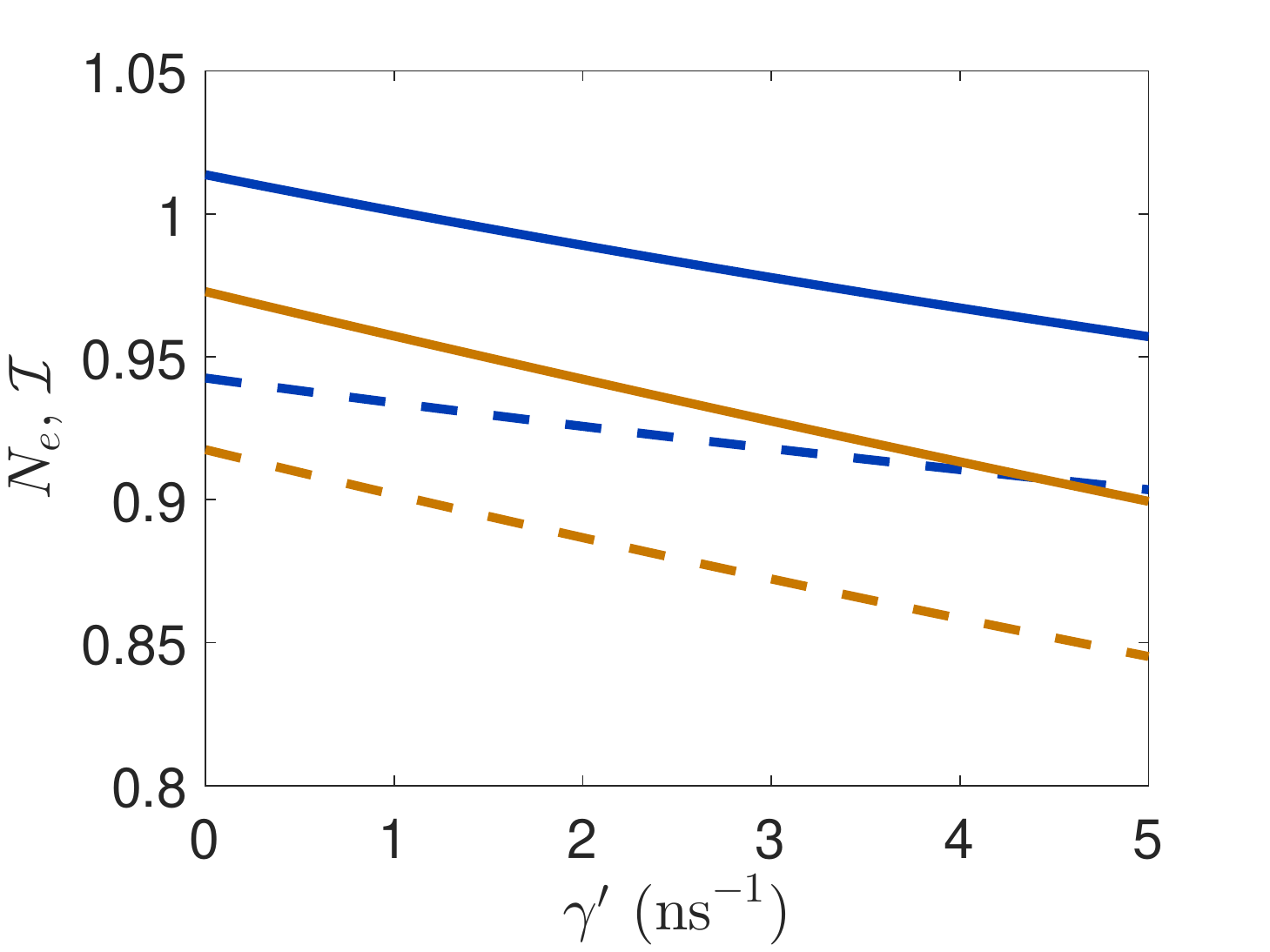}
\caption{\small Emitted photon number $N_e$ (blue) and indistinguishability $\mathcal{I}$ (brown) as a function of pure dephasing $\gamma'$ with phonons (dashed lines) and without (solid lines).}
\label{dephasing_fig} 
\end{figure}

In addition to increased exciton-phonon coupling strengths at higher temperatures (which is intrinsic to the QD), other dissipative processes (e.g., charge noise in QDs) have dephasing rates which are often dependent on temperature~\cite{bimberg01}. In Fig.~\ref{tempvar_fig}  we study the device performance over a wide range of potential operating temperatures by comparing the relative effects of phonon coupling versus temperature-dependent background pure dephasing rates. Following the experimental results in Ref.~\onlinecite{ota09}, we employ an empirical linear pure dephasing correlation $\gamma'(T) = \gamma'_0 +  (2.127 \ \text{ns}^{-1}\text{/K})T$ with $\gamma'_0 = 1 \ \text{ns}^{-1}$ and study the special case of resonant excitation ($\Delta = 0$) with pulse parameters as used elsewhere in this section. Notably, even at low temperatures (4 K), intrinsic phonon-coupling has a less detrimental effect on device figures-of-merit than that of increased pure dephasing rates, which are not fundamental limitations and indeed have been shown to be significantly suppressed in recent experiments~\cite{somaschi16}.

In Fig.~\ref{dephasing_fig}, we show the effect of the (constant) pure dephasing rate on emitted photon number and indistinguishability. At a temperature of $T = 5$ K, we see that it should be possible to achieve simultaneous efficiency and indistinguishability of over 90\% for pure dephasing rates under $\gamma' = 1 \ \text{ns}^{-1}$.

\begin{figure}[t]
\centering
\includegraphics[width=1\linewidth]{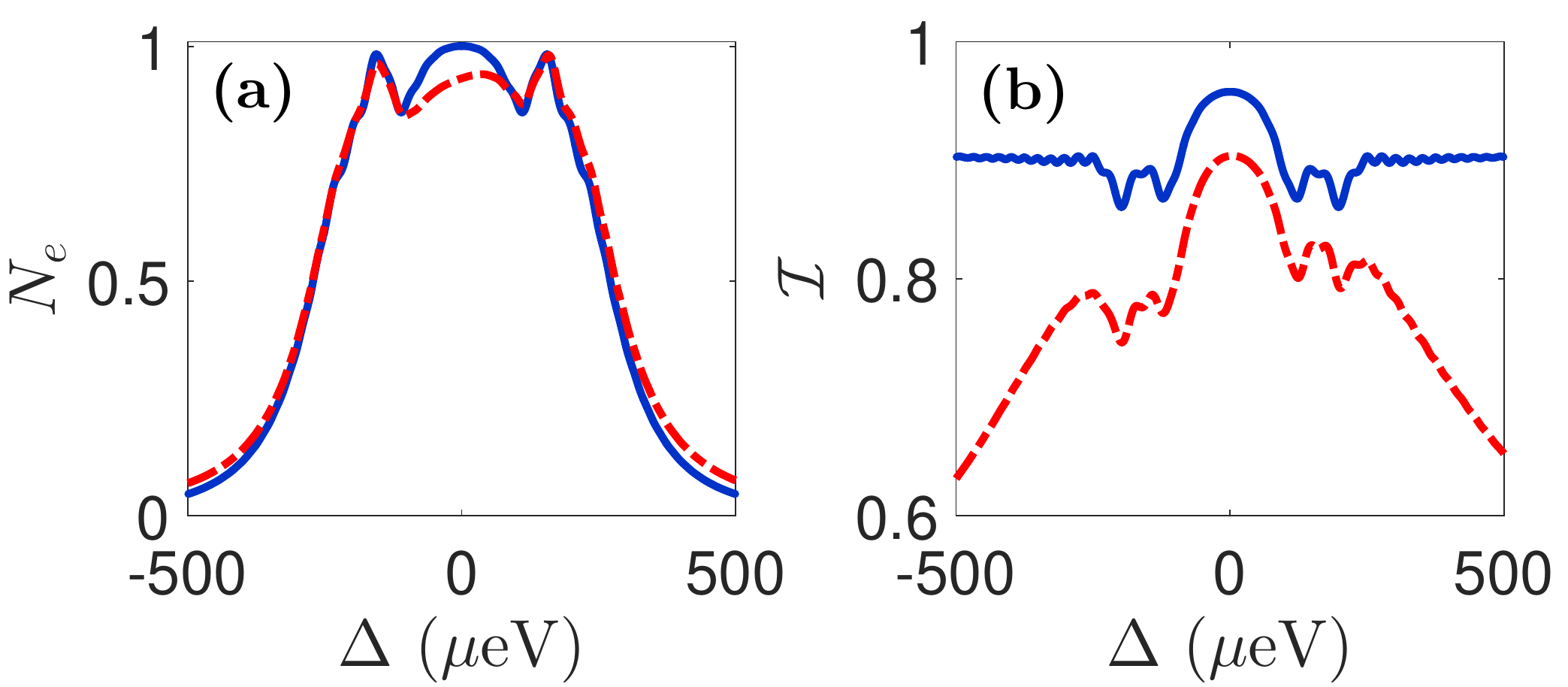}
\caption{\small (a) Emitted cavity photon number and (b) indistinguishability without phonons (solid blue line) and with phonons (dash-dotted red line) as a function of pulse and cavity detuning $\Delta = \Delta_c = \Delta_p$.}
\label{detuning_fig} 
\end{figure}

Next, in Fig.~\ref{detuning_fig}, we study the effects of varying the pump pulse and cavity detuning $\Delta$. Very high efficiencies ($\sim99\%$) are achieved at detunings of $\Delta = \pm \Omega_l = \pm 164.5 \ \mu \text{eV} \  (250 \ \text{ns}^{-1})$, which can be attributed to the Autler-Townes splitting of the biexciton state by  $\pm \Omega_l$ due to the CW laser drive. Positive detunings produce photons of higher indistinguishability, as in this case, the pump pulse and cavity detunings are below resonance with respect to the exciton state transitions, practically avoiding phonon-emission mediated transitions. Since the number of phonons present in a thermal bath is small at low temperatures, phonon-absorption processes are less influential. One potential advantage of off-resonant excitation is that, for positive values of $\Delta$, it ensures that the cavity mode in which photons are emitted into is not only of an orthogonal polarization to the pump and CW light, but also a different frequency, potentially aiding in photon collection and filtering. Even on-resonance, the presence of fine structure anisotropic exchange splitting between of $X$and $Y$ polarized exciton states renders the cavity mode a different frequency than the applied fields, but the degree of the splitting is small ($\sim$10-100 $\mu$eV)  and varies from dot to dot~\cite{lodahl15}. To further study this off-resonant excitation scheme, we take $\Delta = 158 \ \mu \text{eV}$ and show the effect of varying the pulse width in Fig.~\ref{pulsewidth_fig}, along with the corresponding result for resonant excitation ($\Delta = 0$).

\begin{figure}[!t]
\centering
\includegraphics[width=1\linewidth]{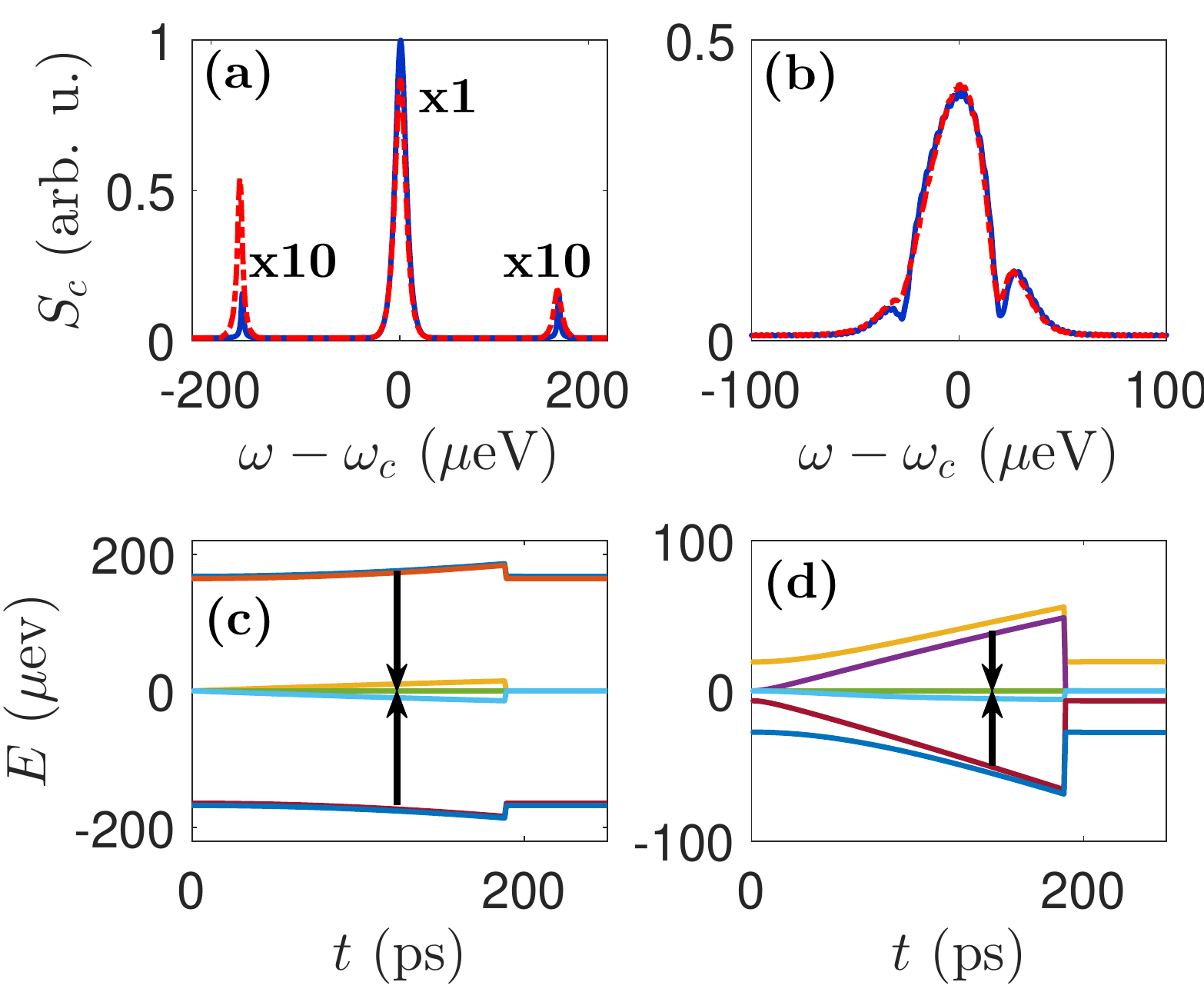}
\caption{\small Top: Cavity-emitted spectra $S_c(\omega)$ for (a) an on-resonance pulse with $\Delta = 0 \ \mu \text{eV}$ and (b) an off-resonance pulse with $\Delta = 158 \ \mu \text{eV}$, with phonons (dash-dotted red line) and without (solid blue line). Bottom: Quasi-eigenergies of system Hamiltonian in rotating frame as a function of time for (c) $\Delta = 0 \ \mu \text{eV}$ and (d) $\Delta = 158 \ \mu \text{eV}$, truncated to a 1 photon (two-dimensional) Fock space, showing non-adiabatic transitions that create sidepeaks in cavity emitted spectrum.}
\label{spectra_fig} 
\end{figure}

It is also interesting to probe the field-induced state dressing that occurs through the cavity-emitted spectrum, which is easily accessed experimentally~\cite{muller07,flagg09,ates09,vamivakas09,ulhaq13}. In Fig. ~\ref{spectra_fig}, we plot the cavity-emitted spectrum with and without phonons for resonant excitation. The emission spectrum from pulsed excitation in the laboratory frame can be found from the Fourier transform of the time-averaged cavity mode first-order correlation function:~\cite{cui06}
\begin{equation}
S_c(\omega) \equiv \text{Re} \Bigg[ \int\limits_{0}^{\infty}d\tau e^{-i(\omega-\omega_c)\tau} \int\limits_{0}^{\infty}dt \langle a^{\dagger}(t+\tau)a(t)\rangle \Bigg].
\end{equation}
The spectrum on-resonance resembles that of the well known Mollow-triplet with a CW drive~\cite{mollow69}, but has a somewhat different physical origin. The sidepeaks in the spectrum arise from biexciton to $Y$-exciton transitions, where the biexciton state is split by $\pm \Omega_l$ due to the presence of the CW drive. For efficient on-resonance STIRAP population transfer, the intermediate (biexciton state) is never significantly populated due to destructive interference in the probability amplitude of transitioning to either of the two energy levels split from the CW laser dressing the $X$-exciton to biexciton transition~\cite{vitanov17}, and the sidepeaks in the cavity-emitted spectrum are thus very small. As the (phonon-induced) dephasing is increased, the intermediate state is populated, leading to off-resonant sidepeaks in the spectrum which are enhanced by phonon absorption and emission processes, asymmetrically favouring phonon emission at low temperatures. Since a lack of sidebands in the emission spectrum (barring any postselection) is a necessary condition (though not sufficient--two-photon emission events must also be suppressed) for indistinguishable photons, this interruption of the STIRAP process by the intermediate eigenstate population can allow for high emitted photon number, but with a cost of lower indistinguishability~\cite{ilessmith16}. Also shown in Fig.~\ref{spectra_fig} are the system time-dependent quasi-eigenenergies corresponding to each respective spectra. In the rotating frame, transitions between different system eigenstates  (shown as black arrows) correspond to side peaks in the cavity-emitted spectrum (the exact location of which depends on the details of the relevant eigenstate populations over time). The eigenvalues of the system Hamiltonian (Eq. \eqref{hsys}) in the rotating frame also can be used to study the mechanism behind the single-photon emission. In ideal STIRAP, the system remains in a zero-eigenvalue state during the entire population transfer process~\cite{vitanov17}, and sidepeaks in the cavity spectrum are very small. For the off-resonant case (Fig.~\ref{spectra_fig} (b)), we see much larger sidepeaks relative to the main peak both with and without phonons, indicating high population of the undesired non-zero eigenvalue dressed states, suggesting that additional phonon-induced dephasing is the dominant decoherence mechanism for this detuning.

Finally, we note that the STIRAP excitation scheme in this work can be modified by eliminating the CW laser coupling the $X$-exciton to biexciton transition and using the pump pulse to couple the ground state to biexciton transition directly via two-photon resonance. This makes the pump process simpler, but  is less effective, requiring longer pulse widths to compensate for the lack of strong Autler-Townes splitting by the CW drive, which increases the influence of dephasing on the system~\cite{vitanov17} and yields indistinguishabilities not higher than 80\% for parameters (excluding the pulse strength and width) similar to those used in the rest of this section.

\section{Conclusions}
\label{conclusions}
In summary, we have analyzed a coherently triggered QD single-photon source utilizing STIRAP with a polaron ME approach which accurately incorporates effects of electron-phonon scattering. Our results, using realistic experimental parameters, show that simultaneous achievement of over 90\% efficiency and indistinguishability, or near-unity efficiency and over 80\% indistinguishability using this set-up should be possible, even in the presence of phonons.

While this source is advantageous in that it combines high indistinguishabilities and efficiencies with easily filtered emitted photons, for near-unity indistinguishabilities (which are typically required in proposals for all-optical quantum computing~\cite{knill05}), recent experiments using resonant pulsed excitation~\cite{somaschi16} and rapid adiabatic passage~\cite{wei14} to invert the exciton state coupled to a cavity have demonstrated higher success than our results would suggest for the STIRAP scheme studied in this work; however, the overall fidelity of these single-photon sources is limited by the fact that simultaneously very high efficiency (brightness) and indistinguishability has yet to be achieved, partially due to the difficulty in effectively filtering and collecting photons under these coherent excitation methods, as the pump field is often resonant with the emitted single photons, requiring polarization discrimination (thus reducing the efficiency and ``on-demand'' nature of the single photon gun). In contrast, by emitting single-photons into a strongly-coupled cavity mode of an orthogonal polarization (and potentially different resonant frequency) to the applied fields, the STIRAP set-up allows for high emission efficiencies and trivially filtered laser light.  \\

\acknowledgments
This work was supported by the Natural Sciences and 
Engineering Research Council of Canada (NSERC) and Queen's University.
We thank Ross Manson and Kaushik Roy-Choudhury for useful discussions.

\appendix*
\section{Rotating frame transformation}
Starting from the Hamiltonian:
\begin{align}\label{dfour}
H =&\  \sum\limits_{\mathclap{S=\{X,Y,XX\}}}\hbar\omega_S\ket{S}\bra{S} + \hbar\omega_{c}a^{\dagger}a + 2\hbar\Big(\Omega_l\cos{(\omega_lt)}\ket{XX}\bra{X} \nonumber \\  + & \ \Omega_p(t)\cos{(\omega_pt)}\ket{X}\bra{g}  + g\cos{(\omega_c t)}\ket{XX}\bra{Y}a + \text{H.c.}\Big) \nonumber \\ + & \ \sum\limits_{\mathbf{q}}\hbar\omega_{\mathbf{q}}b_{\mathbf{q}}^{\dagger}b_{\mathbf{q}} + \sum\limits_{\mathclap{S=\{X,Y,XX\}}} \ \ket{S}\bra{S}\sum\limits_{\mathbf{q}}\hbar\lambda_{\mathbf{q}}^{S}(b_{\mathbf{q}}^{\dagger}+b_{\mathbf{q}}) ,
\end{align}
we move into the interaction picture with the Hamiltonian $\widetilde{H}_I = e^{iH_0 t/\hbar}H_Ie^{-iH_0t/\hbar}$ by defining $H = H_0 + H_{I}$, with 
\begin{align}
H_0 = &\ \hbar\omega_p\ket{X}\bra{X} + \hbar(\omega_p+\omega_l-\omega_c)\ket{Y}\bra{Y} \nonumber \\ +  &  \hbar(\omega_p+\omega_l)\ket{XX}\bra{XX} + \hbar\omega_ca^{\dagger}a.
\end{align}

Expanding the cosines in Eq. \eqref{dfour} as complex exponentials and dropping terms proportional to $\text{exp}(\pm i 2 \omega t)$ (rotating-wave approximation), one arrives at the rotating frame (interaction picture) Hamiltonian used in Eq. \eqref{ham}.

\bibliography{PRB_PaperDraftn}
\end{document}